\documentclass[showpacs,aps,prl,twocolumn,amsmath,amssymb,epsfig]{revtex4-1}

\usepackage{graphicx}
\usepackage{epsfig}
\usepackage{dcolumn}
\usepackage{bm}
\newcommand     {\beq}[1]         { \begin{equation} #1 \end{equation} }

\begin{document}

\title{Fractal frontiers of bursts and cracks in a fiber bundle model of creep rupture}

\author{Zsuzsa Danku}
\author{Ferenc Kun}
\email{ferenc.kun@science.unideb.hu}
 \affiliation{Department of Theoretical Physics, University of Debrecen,
P.O. Box 5, H-4010 Debrecen, Hungary}
\author{Hans J.\ Herrmann}
\affiliation{Computational Physics IfB, HIF, ETH,
  H\"onggerberg, 8093 Z\"urich, Switzerland}
\affiliation{Departamento de Fisica, Universidade Federal do Ceara,
  60451-970 Fortaleza, Ceara, Brazil}

\begin{abstract}
We investigate the geometrical structure of breaking bursts generated
during the creep rupture of heterogeneous materials. Based on a fiber 
bundle model with localized load sharing we show that bursts are compact geometrical
objects, however, their external frontier has a fractal structure which reflects 
their growth dynamics. The perimeter fractal dimension of bursts proved to have 
the universal value 1.25 independent of the external load and of the amount of disorder
in the system. We conjecture that according to their geometrical features breaking bursts 
fall in the universality class of loop-erased self-avoiding random walks with perimeter 
fractal dimension
5/4 similar to the avalanches of Abelian sand pile models. The fractal dimension
of the growing crack front along which bursts occur proved to increase from 1
to 1.25 as bursts gradually cover the entire front.
 \end{abstract}

\maketitle

\section{Introduction}
The fracture of heterogeneous materials proceeds in bursts generated by newly nucleating cracks
or by intermittent propagation steps of crack fronts 
\cite{sethna_crackling_2001,maloy_local_2006,PhysRevLett.112.115502,rosti_crackling_2009}. 
Measuring acoustic emissions of bursts
the fracture process can be decomposed into a time series of crackling events 
\cite{PhysRevLett.110.088702}. 
The analysis of the statistics of crackling noise provides valuable insight
into the dynamics of fracture making it also possible to design methods to forecast
the imminent catastrophic failure \cite{vasseur_heterogeneity:_2015}.
Recent investigations have revealed that beyond the time evolution of crackling time series,
the temporal dynamics of single bursts encodes also
interesting information about the presence and nature of correlations in the underlying 
stochastic process \cite{laurson_evolution_2013,danku_PhysRevLett.111.084302}. 
Less is known, however, about the spatial evolution of bursts and their geometrical structure.

Individual bursts of breaking have been observed experimentally during the propagation
of a planar crack. In the experiments of Refs.\ \cite{maloy_local_2006,tallakstad_local_2011} 
a weak interface between two sintered plastic plates was created introducing also disorder
in a controlled way by sand blasting the surface of the plates. 
The loading and boundary conditions ensured that a
single crack emerged constrained to a plane. Bursts were identified as sudden local jumps
of the front whose spatial and temporal dynamics could be studied by means of high speed 
imaging. Detailed analysis revealed that bursts are composed of extended clusters 
which are nearly compact objects with an anisotropic shape, i.e.\ they are elongated 
along the front. The scaling exponent of the two side lengths of the bounding box 
of clusters was found to be related to the roughness exponent of the crack front 
\cite{tallakstad_local_2011}. 

The experimental findings on the statistical and geometrical
features of bursts could be explained by the crack line model where the long range 
elastic interaction proved to be essential 
\cite{PhysRevLett.101.045501,laurson_avalanches_2010}.
Another interesting approach to the problem was presented in Refs.\ 
\cite{pradhan_failure_2010,hansen_prl_2013,hansen_crackfront_fronttiers2014},
where the interface between a stiff and a soft solid blocks was discretized in terms
of fibers on a square lattice. Slowly increasing the external load a single growing 
crack with a straight average profile was obtained by introducing a gradient 
for the strength of fibers.  It was shown that when crack propagation is controlled
by gradient percolation the crack frontier is fractal with a perimeter dimension 
consistent with the hull exponent $7/4$ of percolation clusters.

Here we present a theoretical investigation of the geometrical structure of breaking bursts 
driven by the short range redistribution of stress. 
We consider a fiber bundle model (FBM) of creep rupture where the localized
load sharing following the breaking of fibers leads to the emergence of a single propagating
crack. In the model
the system evolves under a constant external load, where sudden bursts are triggered along 
a propagating front by slow damaging. The model is well suited to study single bursts because 
creep does not affect the dynamics of breaking avalanches, however, it allows for significantly 
larger avalanche sizes compared to fracture processes occurring under a quasi-statically
increasing external load in FBMs. We show by means of computer simulations that due to the 
short range interaction, bursts are fully connected compact geometrical object with a nearly 
isotropic shape. However, the external frontier of bursts proved to be fractal with a
dimension independent of the load and of the degree of disorder. 
We argue that avalanches driven by the strongly localized redistribution of load
fall in the universality class of loop erased self-avoiding random walks
with the perimeter dimension $5/4$.

\section{Bursts in a fiber bundle model of creep rupture}
To investigate the geometry of bursts we use a fiber bundle 
model which has been introduced recently for the creep rupture
of heterogeneous materials \cite{1742-5468-2009-01-P01021,kun_universality_2008,PhysRevE.85.016116}.
The sample is represented by a parallel set of fibers which are organized on a square
lattice of side length $L$. The fibers have linearly elastic behavior with a
constant Young modulus $E$. Subjecting the bundle to a constant load $\sigma_0$ 
below the fracture strength $\sigma_c$ of the system, the fibers break due to two 
physical mechanisms: immediate breaking occurs when the local load $\sigma_i$ 
on fibers exceeds their fracture strength 
$\sigma_{th}^{i}$, $i=1,\ldots , N$, where $N=L^2$ is the number of fibers. 
Under a sub-critical load $\sigma_0<\sigma_c$ this breaking mechanism would 
lead to a partially failed configuration with an infinite lifetime.
Time dependence arises such that those fibers, which remained intact
under a given load, undergo an aging process accumulating damage $c(t)$. 
We assume that the damage accumulation $\Delta c_i$ per time-step $\Delta t$
has a power law dependence on the local load 
\begin{eqnarray}
\Delta c_i = a\sigma_i^{\gamma}\Delta t,
\label{eq:damlaw}
\end{eqnarray}
where $a$ is a constant and the exponent $\gamma$ controls the characteristic 
time scale of the aging process with $0\leq \gamma < +\infty$. 
The total amount of damage $c_i(t)$ accumulated up to time $t$ can be obtained
by integrating over the entire loading history of fibers 
$c_i(t)=a\int_0^t\sigma_i(t')^{\gamma}dt'$. 
Fibers can sustain only a finite amount of 
damage so that when $c_i(t)$ exceeds the local damage threshold $c^i_{th}$ the fiber breaks.
After each breaking event the load of the failed fiber gets redistributed over 
the remaining intact ones. We assume localized load sharing, i.e.\ after failure events 
the load of broken fibers is equally redistributed over their intact nearest 
neighbors on the lattice. When a fiber without intact nearest neighbors breaks, 
the extension of the neighborhood is gradually increased in steps of one lattice site, 
until the neighborhood contains 
at least one intact fiber. Here such situation 
mainly occurs inside the last, catastrophic burst 
which does not affect the statistics of the data.

Due to the wide separation of the characteristic time scales between the slow damage 
process and immediate breaking a highly complex time evolution emerges:
damage breakings gradually increase the load on the intact fibers
in their vicinity which in turn can trigger sudden bursts of immediate breaking
\cite{kun_fatigue_2007,kun_universality_2008}. 
Eventually, the time evolution of creep rupture
sets in as a series of sudden bursts separated by quite periods of slow damaging. 

\begin{figure}
\begin{center}
\epsfig{bbllx=60,bblly=0,bburx=420,bbury=490,file=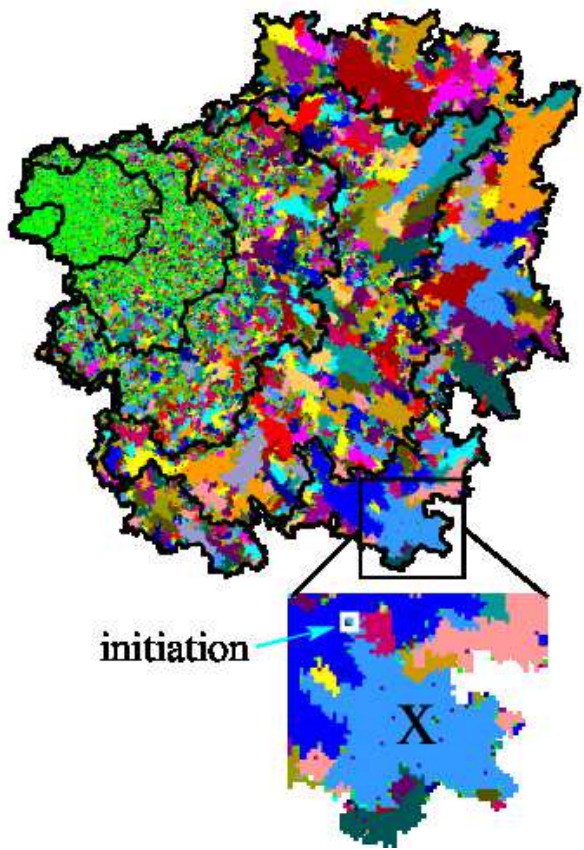, width=8.3cm}
\end{center}
  \caption{(Color online) Evolution of a single crack until macroscopic failure
  of the system at the external load $\sigma_0/\sigma_c=0.01$. 
  The crack starts at the top left corner with a large amount 
  of damage breaking (green area). Damage sequences trigger bursts 
  (spots with randomly assigned colors)
  which occur at the propagating crack front. The bold black lines highlight the 
  position of the crack front at several times during the growth process.
  A magnified view of a small portion of the crack in the black square is presented
  where the burst of light blue color (with a cross in the middle) was initiated 
  by the damage breaking in the white square.
  \label{fig:crack_bursts}
}
\end{figure}
Heterogeneity has two sources in the model, i.e.\ structural 
disorder of materials is represented by the 
randomness of breaking thresholds $\sigma^{th}_i, c^{th}_i$, $i=1,\ldots , N$ of 
the two breaking modes, while additional disorder is induced by the heterogeneous 
stress field generated by the short ranged load redistribution around failed regions. 
For both threshold values we assume uniform distributions over an interval 
$\left[1-\delta_x, 1+\delta_x\right]$, where $x$ stands for $c_{th}$ and $\sigma_{th}$. 
Tuning the width of the distributions $\delta_x$ one can control 
the amount of threshold disorder in the system. In order to promote the effect of stress
concentration on the overall evolution of cracks a small amount of disorder 
is considered for damage $\delta_{c_{th}}=0.2$ with the exponent $\gamma=5$. 
Simulations were carried out 
on a square lattice of linear size $L=401$ varying 
the amount of threshold disorder $\delta_{\sigma_{th}}$ of immediate breaking in the
range $0.2 \leq \delta_{\sigma_{th}}\leq 0.5$. The model has been successfully applied
to describe the time evolution of damage induced creep rupture 
\cite{kun_fatigue_2007,1742-5468-2009-01-P01021,danku_creep_2013}, 
the statistics of crackling bursts 
\cite{kun_universality_2008,PhysRevE.85.016116,danku_frontiers_2014}, 
and even the average temporal profile of bursts \cite{danku_PhysRevLett.111.084302}.
In the present study we focus on the geometrical features of bursts and of the 
crack front.

\section{Structure of single bursts}

The low damage disorder and localized load sharing ensure that at any sub-critical
external load a single growing crack emerges which gradually evolves through bursts 
leading to global failure of the system in a finite time. 
Figure  \ref{fig:crack_bursts} presents a representative
example of a growing crack where individual avalanches of immediate breakings 
can also be distinguished. It can be seen that early stages of the crack growth are
dominated by damage induced breaking of fibers indicated by the large green area. 
However, as the crack reaches larger sizes the high load accumulated along the front
leads to triggering larger and larger bursts (spots of randomly assigned color).
Bursts are characterized by their size $\Delta$ which is the number of fibers 
failing in the cascade of immediate breakings.
Bursts are always induced by damage sequences, however, in later stages of the evolution
a few or a even a single damage breaking can be sufficient to trigger large bursts.
The green dots scattered along the perimeter of extended spots of other colors correspond
to these damage sequences in the figure.
Note that the crack propagation does not have a preferred direction, since no 
gradient is imposed on stress, strain, or strength of fibers.
Hence, the average front position does not
follow a straight line, instead it gets curved as the crack reaches larger sizes.
\begin{figure}
\begin{center}
\epsfig{bbllx=25,bblly=500,bburx=700,bbury=780,file=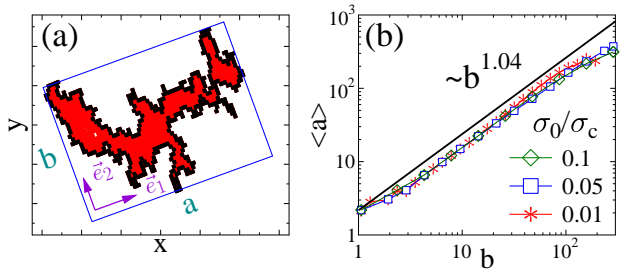, 
width=8.3cm}
\end{center}
  \caption{(Color online)  $(a)$ Definition of the bounding box of a burst. The burst 
  perimeter, highlighted by a bold black line, is formed by those sites of the broken
  cluster which have less then four broken neighbors. $(b)$ Longer side of the 
bounding box $a$ as a function of the shorter one $b$ in a log-log plot.  
  \label{fig:boundingbox}
}
\end{figure}

It can be observed in Fig.\ \ref{fig:crack_bursts} 
that from a geometrical point of view
single bursts are compact objects, i.e.\ the localized load sharing ensures that 
they are dense and they practically do not contain islands 
of intact fibers. However, due to the interplay of the disordered strength
and of the heterogeneous stress field, bursts have a complex external frontier.
In order to characterize the overall geometry of bursts we determined 
the eigenvectors $\vec{e}_1$, $\vec{e}_2$ of the
tensor of inertia of single bursts. Then we constructed the bounding box of bursts as
the rectangle along the vectors $\vec{e}_1$, $\vec{e}_2$ fully surrounding the cluster
as it is illustrated in Fig.\ \ref{fig:boundingbox}$(a)$ where
the longer and shorter edges are denoted by $a$ and $b$, respectively.
For the quantitative characterization of the shape of 
bursts we evaluated the average length of the longer edge $\left<a\right>$ as a function
of the shorter one $b$ which is presented in Fig.\ \ref{fig:boundingbox}$(b)$
for three different load values $\sigma_0/\sigma_c$.
A power law relation of slope very close to unity is obtained showing that 
the side length are simply proportional to each other and the aspect ratio 
$a/b$ does not depend on the burst size $\Delta$.
When long range elastic interaction is taken into account bursts get elongated along the 
front resulting in anisotropic shapes \cite{laurson_avalanches_2010,
hansen_prl_2013,hansen_crackfront_fronttiers2014}.

\begin{figure}
\begin{center}
\epsfig{bbllx=20,bblly=10,bburx=730,bbury=330,file=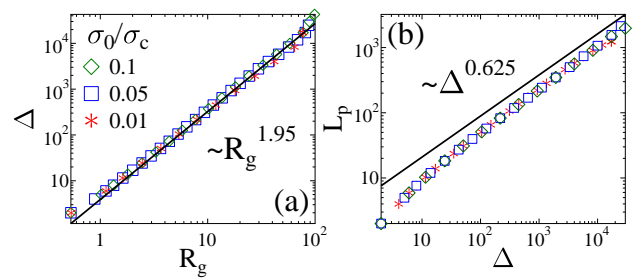, width=8.3cm}
\end{center}
  \caption{(Color online)  $(a)$ Burst size $\Delta$ as a function of the radius of gyration 
  $R_g$ for three values of the external load.  $(b)$ Perimeter of bursts as a function 
  of their area $\Delta$.
  \label{fig:fractal}
}
\end{figure}

Based on the nearly isotropic shape, we determined the radius of gyration $R_g$
of bursts 
\beq{
R_g^2 = \frac{1}{N}\sum_{i=1}^{\Delta}(\vec{r}_i-\vec{r}_c)^2, 
}
where $\vec{r}_i$ ($i=1,\ldots , N$) denotes the position of broken fibers
of the burst and $\vec{r}_c$ is the center of mass of the cluster. 
The value of $R_g$ was averaged over bursts of equal size $\Delta$ accumulating 
all events up to failure. In Fig.\ \ref{fig:fractal}$(a)$ the burst size $\Delta$, i.e.\
the area of the cluster of broken fibers, is plotted as a function of $R_g$ 
where a power law functional form is obtained
\begin{eqnarray}
\Delta \sim R_g^{D}.
\end{eqnarray} 
The value of the exponent is $D\approx 2$ which shows the high degree 
of compactness of bursts. Note that the statistics of the data is mainly 
determined by the size distribution of bursts. Recently, we have shown that 
the size of bursts $\Delta$ is power law distributed with an exponential 
cutoff controlled by the external load 
\cite{kun_universality_2008,danku_PhysRevLett.111.084302}.
Fluctuations are kept low in the figures by simulating an ensemble of 10000 samples 
for each 
parameter set. The $\Delta(R_g)$ curves obtained at different loads
$\sigma_0$ fall on the top of each other, practically the only effect of $\sigma_0$
is that it controls the upper cutoff of $\Delta$ since at lower loads smaller 
bursts are triggered \cite{PhysRevE.85.016116,danku_PhysRevLett.111.084302}. 

The most remarkable feature of avalanches is that they have a tortuous frontier
composed of a large number of valleys and hills (see Fig.\ \ref{fig:crack_bursts} and 
Fig.\ \ref{fig:boundingbox}). 
The structure of the burst perimeter is determined by the interplay of the disorder 
of the failure thresholds $\sigma_{th}$ and of the stress field during the growth
of the avalanche: The avalanche usually starts from a single breaking fiber. 
As the load gets redistributed some of the 
surrounding intact fibers may exceed their failure threshold and break leading to
an expansion of the burst. Bursts proceed through such subsequent breaking and load 
redistribution steps until all intact fibers along the burst frontier 
can sustain the elevated load. 
For a quantitative assessment we determined the number of perimeter 
sites $L_p$ of each burst as a function of the burst size $\Delta$. 
On the square lattice those broken fibers are identified as perimeter sites
which have less than four broken neighbors in the same cluster. 
Inside bursts a very small amount of intact fibers may remain isolated, they 
are removed before the perimeter identification.
In Fig.\ \ref{fig:boundingbox}$(a)$ the perimeter is highlighted by a bold black line
while the bulk of the cluster is red.

\begin{figure}
\begin{center}
\epsfig{bbllx=40,bblly=20,bburx=380,bbury=330,file=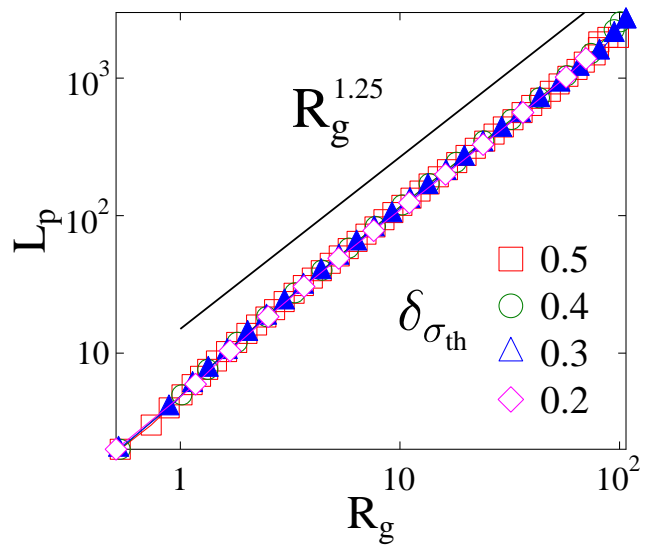, width=8.3cm}
\end{center}
  \caption{(Color online) Perimeter length $L_p$ of bursts as a function 
  of the radius of gyration $R_g$ for four different amounts
  of threshold disorder $\delta_{\sigma_{th}}$. All the curves fall on 
  top of each other.
  \label{fig:fractaldisorder}
}
\end{figure}

The perimeter length $L_p$ averaged over all bursts of a fixed size $\Delta$ is presented
in Fig.\ \ref{fig:fractal}$(b)$ as a function of $\Delta$.
In the regime of large burst sizes a power law is evidenced 
\begin{eqnarray}
L_p \sim \Delta^{\xi},
\end{eqnarray}
where the value of the exponent $\xi$ was obtained as $\xi = 0.625\pm 0.015$.
This feature implies that the avalanche frontier is a fractal with fractal
dimension $D_f=\xi D=1.25\pm0.03$, where $D=2$ was substituted. 
It is important to emphasize that the fractal dimension $D_f$ proved to be universal, i.e.\
it neither depends on the external load $\sigma_0$ nor on any details of damage accumulation
such as $\gamma$, as long as single crack propagation is ensured in a 
heterogeneous environment. The only role of the damage mechanism is that fibers 
breaking due to damage initiate the bursts. Once the burst has started it is driven 
by the gradual redistribution of load. 
Figure \ref{fig:fractaldisorder} presents the perimeter length of avalanches as a 
function of the radius of gyration for several different values of the width 
$\delta_{\sigma_{th}}$ of the threshold distribution of immediate breaking. 
It can be observed that the curves fall on top of each other, which indicates
that the amount of disorder does not have a noticeable effect until its value is sufficiently
high. The structure of the external frontier of bursts emerges as an outcome
of the interplay of the short range stress redistribution and of the strength 
disorder of fibers.

It is interesting to note that avalanches of other type of systems with short range interaction
show a striking similarity to the bursts of our LLS FBM. The Abelian Sandpile Model (ASM)
is a paradigmatic model of self organized criticality 
\cite{majumdar_PhysRevLett.68.2329,dhar_PhysRevLett.64.1613} 
where topling (overstressed) lattice sites relax by redistributing sand grains over their local
neighborhood. Majumdar showed that the frontiers of avalanches driven by short range redistribution 
in ASM can exactly be mapped to Loop Erased Random Walks (LERW) in two dimensions, 
and determined their fractal dimension exactly $D_f=5/4$ \cite{majumdar_PhysRevLett.68.2329}.  
Based on the growth dynamics, geometrical features, and robustness of the fractal dimension
of burst perimeters we conjecture that breaking avalanches of 
FBMs fall in the universality class of
LERWs similarly to avalanches of granular piles \cite{majumdar_PhysRevLett.68.2329}. 

\section{Structure of the propagating front}
Burst are always initiated along the propagating crack front where the load 
redistribution after damage breakings can give rise to immediate breaking of fibers. 
It can be observed in Fig.\ \ref{fig:crack_bursts} that the early stage 
of crack growth is dominated by the damage mechanism, the crack size has to reach a 
threshold size to achieve a sufficiently high stress concentration at the 
crack frontier to trigger bursts. As the crack grows all the load kept by the fibers 
inside the crack area accumulates at the crack front which gives rise to 
larger and larger bursts. The localized load sharing ensures that the crack is a 
dense set of broken fibers and there are no broken clusters ahead of the
front (no process zone can form). It can be observed in Fig.\ \ref{fig:crack_bursts}
that when damage breakings dominate at the early stage of crack growth a smooth front
is formed, while it gets tortuous as burst gradually overtake the control of 
propagation. It follows that the key feature which determines the degree of smoothness 
of the front is the ratio between the damage and immediate breakings along the perimeter.
\begin{figure}
\begin{center}
\epsfig{bbllx=50,bblly=50,bburx=350,bbury=350,file=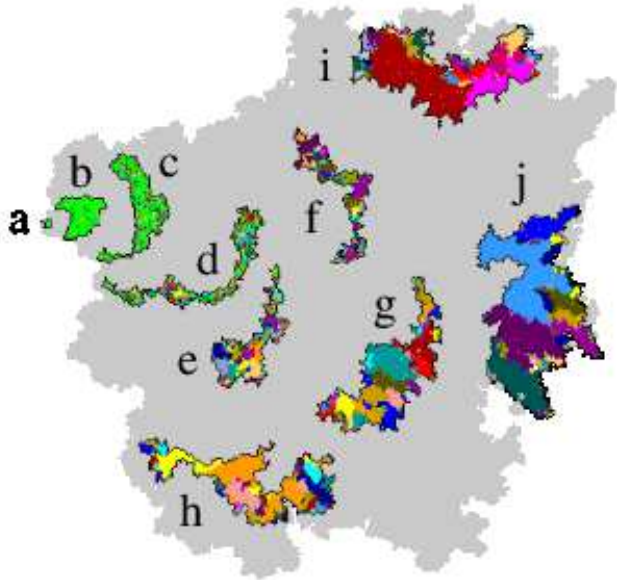, width=8.3cm}
\end{center}
  \caption{(Color online) Clusters of bursts inside the growing crack of Fig.\ 
  \ref{fig:crack_bursts}. The same color coding is used as in Fig.\ \ref{fig:crack_bursts}, i.e.\
  the damage breakings which trigger bursts are in green,
  while the bursts have randomly assigned colors different from green.
  The clusters denoted by $(a), \ldots , (j)$ are generated at different stages of crack growth,
  where $(a)$ indicates the initial damage cluster without any burst. The other clusters
  correspond to different ratios $L_p^d/L_p$ of the perimeter length $L_p$ 
  and of the number of damage breakings in the perimeter $L_p^d$:
  $(b)$ $0.8-0.9$, $(c)$ $0.5-0.6$,$(d)$ $0.4-0.5$,
  $(e)$ $0.2-0.3$,$(f)$ $0.2-0.3$,$(g)$ $0.1-0.2$,$(h)$, $(i)$, $(j)$ $0-0.1$.
  \label{fig:crack_cluster}}
\end{figure}

\begin{figure}
\begin{center}
\epsfig{bbllx=5,bblly=455,bburx=355,bbury=750,file=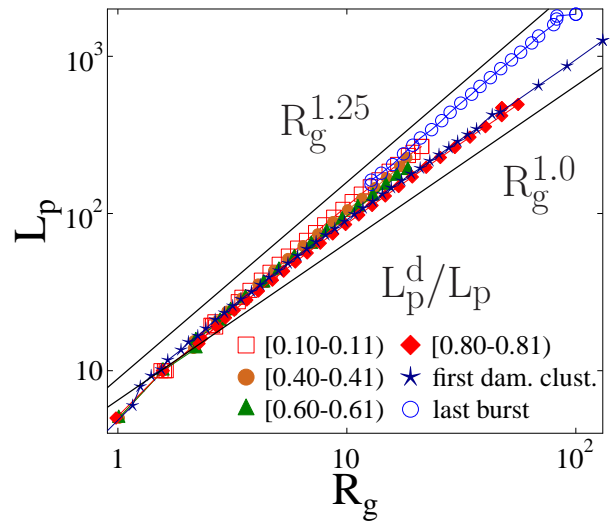, width=8.3cm}
\end{center}
  \caption{(Color online) Perimeter length of clusters of bursts inside the crack for several
  values of the ratio $L_p^d/L_p$ of the amount of damage breakings 
  $L_p^d$ along the cluster perimeter 
  and of the perimeter length $L_p$.
  For completeness, the initial cluster of damage breakings ({\it first dam. clust.}) and the 
  last burst of the crack are also included. Damage breaking favors a smooth cluster boundary 
  which implies effectively $D_f=1$. As the fraction $L_p^d/L_p$ decreases bursts determine 
  the cluster perimeter
  so that a gradual crossover occurs to $D_f=1.25$.
  \label{fig:crossover}
}
\end{figure}

In order to quantify the change of structure of the crack front,
inside the crack we identified clusters of broken fibers formed by the 
burst and by the damage breakings which occurred within certain 
time intervals. Such clusters are highlighted in Fig.\ \ref{fig:crack_cluster}
inside the crack of Fig.\ \ref{fig:crack_bursts}. We determined the total perimeter length 
$L_p$ of the clusters and the number of damage breaking $L_p^d$ contained in $L_p$.
Then we grouped the clusters according to the value of the ratio $L_p^d/L_p\leq 1$ using
$0.01$ for the bin size. 
Figure \ref{fig:crossover} presents the perimeter $L_p$ of clusters for a few selected
bins of $L_p^d/L_p$ as a function of the radius of gyration $R_g$.
It can be observed in the figure that when damage breaking dominates $L_p^d/L_p\approx 1$ 
along the external frontier of clusters, the perimeter fractal dimension takes the effective 
value $D_f=1.0$, as it is expected for smooth lines. However, as the fraction $L_p^d/L_p$ decreases 
immediate breakings of bursts control more the geometry of the perimeter so that a
gradual crossover is obtained to the higher value $D_f=1.25$ of the fractal 
dimension of burst perimeters. 
It follows that in our localized load sharing 
fiber bundle model of creep rupture the fractal dimension of the propagating 
crack front depends on at which stage of propagation it is measured. 

\section{Discussion}
We investigated the geometrical structure of breaking bursts and of the crack front
in the framework of a fiber bundle model with localized load sharing. The model
has been developed for the creep rupture of heterogeneous materials under constant
sub-critical external loads. In the model fibers break due to two reasons:
When the local load surpasses the strength of a fiber immediate breaking occurs. 
Time dependence is introduced by the second breaking mechanism, i.e.\ loaded fibers 
accumulate 
damage and break when the amount of damage exceeds a respective threshold. 
Damaging fibers trigger bursts of immediate breakings which occur along the front 
of a growing crack as sudden local advancements of the front. 
The bursting dynamics is mainly controlled by the localized redistribution of
load after fiber failures (short range interaction) and by the strength disorder.

We showed that due to the localized load sharing single bursts form compact 
geometrical objects of a nearly isotropic shape. However, the external
frontier of bursts has a fractal structure characterized by the perimeter 
fractal dimension $D_f=1.25$. Computer simulations revealed that the value
of the fractal dimension is universal, i.e.\ it does not depend neither 
on the external load nor on the amount of disorder in the range considered.
The main role of the external load is that it controls the characteristic 
time scale of the failure process and the number and cutoff
size of bursts that occur up to macroscopic failure, while 
the precise amount of disorder does not have a crucial effect until it is
sufficiently high to prevent sudden early collapse (brittle failure)
of the system. 

Based on the value of the perimeter fractal dimension and on its striking 
universality, we conjecture that crackling bursts controlled by the localized 
redistribution of load fall in the universality class of loop erased self avoiding
random walks similarly to the avalanches of Abelian sand pile models 
\cite{majumdar_PhysRevLett.68.2329,dhar_PhysRevLett.64.1613}. When
long range elastic interaction along the front dominates the crack propagation 
different geometrical features were found \cite{tallakstad_local_2011}.

Our simulations showed that the geometrical structure of the crack front changes 
as the system evolves: damage breakings favor a smooth front while the bursts tend
to make it more tortuous. The key parameter to control the front geometry proved
to be the fraction of damage breakings along the front. For high values
of the damage fraction, corresponding to the early stage of crack growth,
the fractal dimension tends to 1, while for low fractions, obtained at larger crack sizes,
a gradual crossover occurs to the burst perimeter dimension 1.25.

Compared to simple fiber bundle models of fracture with quasi-static loading, the creep 
dynamics has the advantage that a single growing crack emerges under a constant 
external load and large bursts develop along the crack front. 
However, since no gradient is imposed on stress, strain, or strength of fibers
the crack front is curved. This implies that no average front position can be defined
so that we cannot study the roughness of the front.

In our analysis the parameters of the damage mechanism, i.e.\ the $\gamma$ exponent
and the amount of disorder of the damage thresholds $\delta_{c_{th}}$ were set 
to ensure the dominance of stress concentration in the failure process which leads
to the emergence of a single growing crack. However, for low exponent $\gamma \to 0$
and high disorder $\delta_{c_{th}}\to 1$ the breaking process gets disorder dominated 
where a large number of cracks grow simultaneously \cite{PhysRevE.85.016116}.  
We note that the geometrical structure of bursts does not depend on the precise 
value of the damage parameters, however, from numerical point of view it is advantageous
to perform the analysis in the phase of single crack growth because here bursts and crack 
reach larger sizes.

\begin{acknowledgments}
We thank the projects TAMOP-4.2.2.A-11/1/KONV-2012-0036, 
TAMOP-4.2.2/B-10/1-2010-0024, OTKA K84157, and ERANET\_HU\_09-1-2011-0002 
and ERC Advanced Grant 319968 FlowCCS.
\end{acknowledgments}

\bibliography{/home/feri/papers/statphys_fracture}

\end{document}